\documentclass[12pt,graphicx]{article}

\usepackage{amssymb}
\usepackage{graphicx}

\begin{document}

\title{Asymmetric Anomalous Diffusion:
an Efficient Way to Detect Memory in Time Series}
\author{Paolo Grigolini$^{1,2,3}$, Luigi Palatella$^{1}$, Giacomo Raffaelli$^{1}$}
\date{}
\maketitle

\begin{center}
{\em $^{1}$Dipartimento di Fisica dell'Universit\`{a} di Pisa and
INFM \\ Piazza Torricelli 2, 56127 Pisa, Italy \\
$^{2}$Istituto di Biofisica del Consiglio Nazionale delle
Ricerche,\\ Area della Ricerca di Pisa, Via Alfieri 1, San Cataldo,
56010,Ghezzano-Pisa, Italy \\
$^{3}$Center for Nonlinear Science, University of North Texas,
P.O. Box 311427, Denton, Texas 76203-1427 \\}
\end{center}


\begin{abstract}
{We study time series concerning rare events. The occurrence of a
rare event is depicted as a jump of constant intensity  always
occurring in the same direction, thereby generating an asymmetric 
diffusion process. We consider the case where the
waiting time distribution is an inverse power law with index $\mu$.
We focus our attention on $\mu <3$, and we evaluate the scaling
$\delta$ of the
resulting diffusion process. We  prove that $\delta$ gets
its maximum value, $\delta = 1$, corresponding to the ballistic
motion, at $\mu =2$. We study the resulting diffusion process by
means of joint use of the continuous time random walk and of the
generalized central limit theorem, as well as adopting a numerical
treatment.
We show that rendering asymmetric the diffusion process yields the
significant benefit of enhancing the value of the scaling parameter
$\delta$. Furthermore, this scaling parameter becomes sensitive to the
power index $\mu$ in the whole
region $1<\mu<3$. Finally, we show our method in action on real data
concerning human heartbeat sequences.}
\end{abstract}


\section {Introduction}

One of the basic expectations of the statistical analysis of time series of
biological, sociological and financial interest\cite{eugene} is that
the process under study might depart from fully random behavior.
Following the prescriptions of the authors
of Refs.\cite{detrendingindna,detrendinginheartbeating}
we use the time series to create a sort of random walk trajectory
whose statistical properties are then carefully studied to detect the
deviations from ordinary Brownian motion. 
To do so a given site of
the time series is interpreted as the ``time'' at which the walker
makes a jump. The intensity of the jump is determined by the
value of the time series at that given site, and the walker jumps
either forward or backward according to whether the sign of the value
is positive or negative. 
The most natural way to establish a
comparison between this walker and the ordinary random walker of
statistical mechanics would be to compare the two resulting diffusion
processes to one another, but this would imply that infinitely
identical walkers are available. Unfortunately a time series means
only one trajectory, thereby generating the problem of how to create a
number of ``independent'' trajectories large enough to produce a
diffusion process. The authors of
Refs.\cite{detrendingindna,detrendinginheartbeating} addressed this
problem with an ingenious technique called \emph{Detrended 
Fluctuation Analysis} (DFA)
which allows us to bypass this difficulty. 
The whole pseudo-random 
trajectory is divided
into non-overlapping intervals of length $l$, 
each of which can be 
imagined as being a
trajectory running for a "time" $l$. 
Thus we have a collection of a 
relatively large
number of "independent" trajectories that can be used to produce the 
wanted diffusion process.
These trajectories might be affected by a local trend and this local 
trend might result in an
artificial deviation from ordinary Brownian motion. For this reasons
the authors of Refs.\cite{detrendingindna,detrendinginheartbeating}
for any interval evaluate the local bias and then record only
the deviation from the local trend. The square root of the resulting variance,
denoted by them with the symbol $F(l)$, is expected to yield for very 
large $l$'s
\begin{equation}
F(l)  \propto l^{\delta},
\label{tenetofdfa}
\end{equation}
where $\delta$ denotes the scaling parameter, a crucial property to 
assess the departure
from ordinary Brownian motion. In fact, ordinary Brownian motion 
should yield $\delta_{B} = 0.5$
and the deviation of the observed value of $\delta$ from $\delta_{B} 
= 0.5$ is considered to be a
measure of the deviation from a totally random behavior.

It has to be pointed out that scaling is a property implying that
\begin{equation}
p(x,l) = \frac{1}{l^{\delta}}F(\frac{x}{l^{\delta}})
\label{rigorousdefinition}
\end{equation}
and that in some special cases, discussed in this paper, the scaling 
value detected by means
of Eq.(\ref{tenetofdfa}) does not afford a satisfactory account of the property
of Eq.(\ref{rigorousdefinition}).
Furthermore, as we shall see, the scaling $\delta$, even if it 
reflects correctly
the property of Eq.(\ref{rigorousdefinition}), might not disclose an 
exhaustive information
on the dynamic source of the deviation from ordinary Brownian motion.

Here we illustrate a technique of analysis, called Diffusion Entropy
(DE) method, originally applied to detect memory effects in
time series of sociological interest\cite{nicola}. We shall show that 
this technique can,
in part, solve the earlier mentioned problems.
We use a sequence of the same
kind as that adopted in earlier work to mimic the statistical properties
of DNA sequences\cite{allegrodna}.  This sequence consists of
strings of extended length, filled with only $+$'s or $-$'s. We
consider as significant event only the change of sign. Thus, this kind
of sequence is changed into a new sequence realized by assigning
the vanishing value to the sites with a sign identical to that of the
following one.
Only the sites with a sign
opposite to that of the next site are assigned a non vanishing value,
which is always the same positive quantity, $W$. 
Thus we obtain a sequence of rare events. When one of these events
occurs,  the
random walker makes a jump of constant intensity, always in the same direction.
We prove that the resulting diffusion process makes much more
ostensible the correlated nature of the original time series. 

We also show that this procedure makes faster the detection of the
proper scaling parameter $\delta$, namely, a scaling value which is a
fair reflection of the property of Eq.(\ref{rigorousdefinition}).
As earlier mentioned, the
methods determining scaling through variances might yield scaling
parameters significantly departing from the ideal scaling value. This
is the scaling that an artificial time series, whose asymptotic
behavior is dictated by compelling theoretical arguments, is expected
to reach in the long-time limit.

The outline of the paper is as follows.
In Section \ref{II} we describe the
main properties of the DE method,
and  with the help of an
artificial sequence we show that the observation
of a symmetric diffusion process results
in a slow convergence towards the emergence of
the expected scaling parameter. Section \ref{III}
illustrates a theoretical approach to the
scaling of the asymmetric diffusion process using
two distinct procedures, the Continuous Time Random Walk (CTRW)
and the Generalized Central Limit Theorem (GCLT), yielding
the same result. This section also shows, with the help of
the DE method applied to some artificial sequences, 
that the detection of the
correct scaling is faster than in the case of symmetric
diffusion. Finally, Section \ref{cuore} shows our detection
technique at work on real data. The illustrative example of
Section \ref{cuore} is based on
a time series resulting from the record of human heart beats.
A balance on the results obtained in this paper is made in Section \ref{V}.

\section{Diffusion Entropy}\label{II}

The main idea of this approach to scaling is remarkably simple.
Let us consider a sequence of $M$ numbers $ \xi_{i}$ , with  $i = 
1,  \ldots , M$.
The purpose of the DE algorithm is to establish the possible
existence of a scaling, either normal or anomalous, in the most
efficient way as possible, without altering the data with any form
of detrending. Let us select first of all an integer number
$l$, fitting the condition $1 \leq l \leq M$.
This integer number will be referred to by us as ``time''. 
For any given time $l$ we can find $M - l +1$ sub-sequences defined by
\begin{equation}
\xi_{i}^{(s)} \equiv \xi_{i + s}, \quad   \quad s = 0, 
\ldots ,  M-l.
\label{multiplicationofsequence}
\end{equation}
For any of these sub-sequences we build up a diffusion
trajectory, labelled with the index
$s$, defined by the position
\begin{equation}
x^{(s)}(l) = \sum_{i = 1}^{l} \xi_{i}^{(s)}
= \sum_{i = 1}^{l} \xi_{i+s}.
\label{positions}
\end{equation}

Let us imagine this position as referring to a Brownian 
particle that at regular intervals of time has been jumping forward or
backward according to the prescription of the corresponding
sub-sequence of Eq.(\ref{multiplicationofsequence}). This means
that the particle before reaching the position that it holds at
time $l$ has been making $l$ jumps. The jump made at the
$i$-th step has the intensity $|\xi_{i}^{(s)}|$ and is forward or
backward according to whether the number $\xi_{i}^{(s)}$ is
positive or negative.

We are now ready to evaluate the entropy of this diffusion process.
To do that we have to partition the $x$-axis into cells of size
$\epsilon(l)$. When this partition is made we have to label the
cells. We count how many particles are found in the same cell at a
given time $l$. We denote this number by $N_{i}(l)$. Then
we use this number to determine the probability that a particle
can be found in the $i$-th cell at time $l$, $p_{i}(l)$, by means of
\begin{equation}
p_{i}(l) \equiv  \frac{N_{i}(l)}{(M-l+1)} .
\label{probability}
\end{equation}
At this stage the entropy of the diffusion process at the time $l$
is determined and reads
\begin{equation}
S_{d}(l) = - \sum_{i} p_{i}(l) ln [p_{i}(l)].
\label{entropy}
\end{equation}
The easiest way to proceed with the
choice of the cell size, $\epsilon(l)$, is to assume it independent
of $l$ and determined by a suitable fraction of the square root of
the variance of the
fluctuation $\xi(i)$.

Before proceeding with the illustration of how the DE method works,
it is worth making a comment on the way we use to define the trajectories.
The method we are adopting is based on the idea of a moving window of
size $l$ that
makes the $s-th$ trajectory closely correlated to the next, the
$(s+1)-th$ trajectory. The two trajectories have $l-1$ values in
common. The DFA, on the contrary, is based on non-overlapping windows,
and consequently trajectories with different labels are totally
independent the one from the other.
A motivation for our choice is that we are having in mind a possible
connection with the Kolmogorov Sinai (KS) entropy\cite{beck,dorfman}.
The KS entropy of a symbolic sequence is evaluated by moving a window
of size $l$ along the sequence. Any window position corresponds to a
given combination of symbols, and, from the frequency of each
combination, it is possible to derive the Shannon entropy $S(l)$. The
KS entropy is given by the asymptotic
limit $lim_{l \rightarrow \infty} S(l)/l$. We believe that the same
sequence, analyzed with the DE method, at the large values of $l$ 
where $S(l)/l$ approaches the KS value,
 must yield a well defined scaling $\delta$. To realize this
correspondence we carry out the determination of the DE using the
same criterion of overlapping windows as that behind the KS entropy.

Details on how to deal with the transition from the
short-time regime, sensitive to the discrete nature of the
process under study, to the long-time limit where both space
an time can be perceived as continuous, are given in
Ref.\cite{nicvit}. Here we make the simplifying assumption
of considering so large times as to make the continuous assumption
valid.
In this case the trajectories, built up with the above illustrated procedure,
correspond to the following equation of motion:

\begin{equation}
\frac{dx}{dt} = \xi(t) ,
\label{equationofmotion}
\end{equation}
where $\xi(t)$ denotes the value that the time series under study
gets at the $\l-th$  site. This means
that the function $\xi(l)$ is
depicted as a function of $t$, thought of as a continuous time $t = l$.
In this case the Shannon entropy reads
\begin{equation}
S(t) = - \int_{-\infty}^{\infty} dx \, p(x,t) ln [p(x,t)].
\label{continuousshannonentropy}
\end{equation}
We assume
\begin{equation}
p(x,t) = \frac{1}{t^{\delta(t)}} \, F\left( 
\frac{x}{t^{\delta(t)}}\right).
\label{generalizedscaling}
\end{equation}
This is a generalization of the ordinary scaling assumption of
Eq.(\ref{rigorousdefinition}),  which, in fact, can
be recovered by assuming $\delta(t)$  to be time independent so that
we can denote it as $\delta$.
For the sake of simplicity we make the simplifying assumption  that
$F(y)$ maintains its form, namely, that the  statistics of the
process are unchanged. This is questionable in the case where the
scaling is a function of time. However, making this assumption we
obtain the following results.
Let us plug Eq.(\ref{generalizedscaling}) into
Eq.(\ref{continuousshannonentropy}). After a simple algebra, we get:
\begin{equation}
S(\tau) = A + \delta(\tau) \tau ,
\label{keyrelation}
\end{equation}
where
\begin{equation}
A \equiv -\int_{-\infty}^{\infty} dy \, F(y) \, ln [F(y)]
\label{ainthecontinuouscase}
\end{equation}
and
\begin{equation}
\tau \equiv ln (t) .
\label{logarithmictime}
\end{equation}

It is evident that this kind of technique to detect scaling does
not imply any form of detrending, and this is one of the reasons why
some attention should be devoted to it. An interesting way to
check the efficiency of this technique is realized by the study
of the artificial sequence of Ref. \cite{marcoluigi}. This
sequence is built up in such a way as to realize long sequences
of either $+$'s or $-$'s. The probability of finding a sequence
of only $+$'s or only $-$'s of length $t$ is given by
\begin{equation}
\label{chosenanalyticalform}
\psi(t) = (\mu - 1) \frac{T^{\mu-1}}{(t+T)^{\mu}}.
\end{equation}
Here we focus our attention on the condition $\mu < 3$ and we
raise the reader's attention on the interval $[2,3]$.
In fact, this kind of sequence is the same as that adopted
in earlier work \cite{allegro} for a dynamic derivation of L\'{e}vy
diffusion, which shows up when the
condition $2 < \mu < 3$ applies.
It corresponds to a particle travelling with constant velocity
throughout the whole time interval corresponding to either only +'s
or only -'s, and changing direction with no rest, at the end of any
string with the same symbols.

\begin{figure}[h]
\begin{center}
\includegraphics[scale=.6]{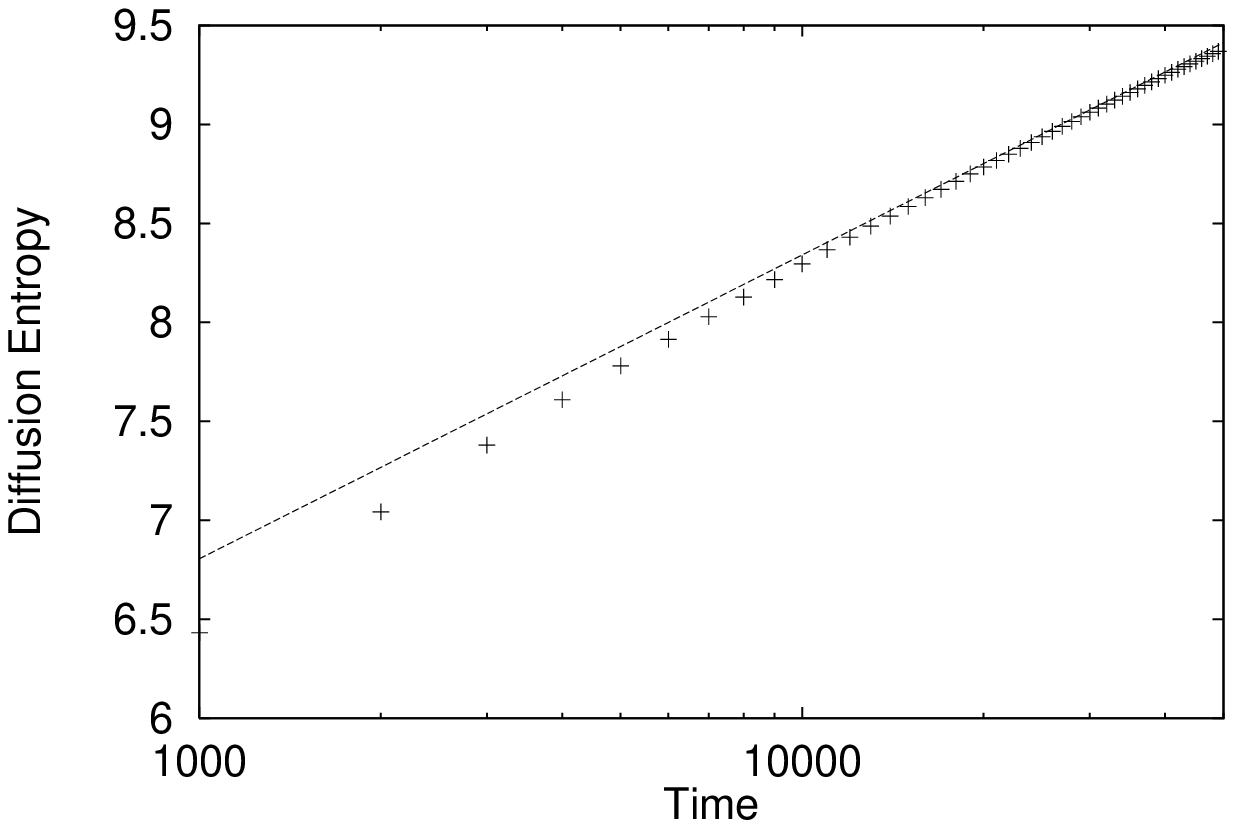}
\caption{\label{SVMDE}
The diffusion entropy as a function of time. The numerical method is
applied to the artificial sequence of Section \ref{II}, with $\mu = 2.5$,
studied according to the SVM prescription. According to the
theoretical arguments of the text the scaling parameter $\delta$ is the
slope of the straight line fitting the numerical results at large
times, which yields in this case $\delta = 2/3 = 1/(\mu -1)$ ($T = 10$).}
\end{center}
\end{figure}

We will refer to this model as {\em Symmetric Velocity Model} (SVM).
We know from the theory of Ref.\cite{allegro} that the scaling
of the resulting diffusion process when $2<\mu<3$ is
\begin{equation}
\delta = \frac{1}{\mu -1}.
\label{levyscaling}
\end{equation}
Note, however, that this diffusion process has a finite propagation
front, with ballistic peaks showing up at both $x = t$ and $x = -t$.
The intensity of these peaks is proportional
to the correlation function
\begin{equation}
\Phi_{\xi}(t) = \left ( \frac{T}{T+t} \right )^{\mu -2}.
\label{correlatonfunction}
\end{equation}
As a consequence of this fact, the whole distribution does not
have a single rescaling. In fact, the distribution enclosed
between the two peaks rescales with $\delta$ of
Eq.(\ref{levyscaling}) while the peaks are associated to $\delta =1$.
Furthermore, it is well known\cite{allegro} that the scaling of the second
moment is given by
\begin{equation}
\delta = \frac{4 - \mu}{2}.
\label{secondmomentscaling}
\end{equation}
Thus, it is expected that the scaling detected by the DE method
might not coincide with the prediction of Eq.(\ref{levyscaling})
for the whole period of time corresponding to the presence of peaks
of significant intensity. We think that the L\'{e}vy scaling of 
Eq.(\ref{levyscaling}) will show up at long times, when the peak
intensity is significantly reduced.
This conjecture seems to be supported
by the numerical results illustrated in Fig.\ref{SVMDE}. We see in fact that
the scaling predicted by Eq.(\ref{levyscaling}) is reached after an extended
transient, of the order of about 20,000 in the scale of Fig.\ref{SVMDE}.
This time interval is about 2000 larger than the value assigned to the
parameter T, of Eq.(\ref{chosenanalyticalform}),
which is, in fact, in the case of Fig.\ref{SVMDE}, T= 10.

In conclusion, this section proves that the DE method applied
to the SVM yields, for the scaling parameter $\delta$, the correct
value of Eq.(\ref{levyscaling}),
rather than the value that would be obtained
measuring the variance of the diffusion process, Eq.(\ref{secondmomentscaling}).
However, the time necessary to make this correct value emerge is very large.
Furthermore, as proved by the theory of Ref.\cite{zumofen}, the adoption of
SVM  would make the scaling parameter $\delta$ insensitive to $\mu$ in
the whole interval [1,2]. This means that the adoption of the DE
method would not allow us
to distinguish a process with $\mu$ very close to 1 from one with $\mu$
very close to 2. In Section \ref{III} we shall see how to overcome these
limitations.

\section{Asymmetric Diffusion}\label{III}

Let us consider again the kind of artificial sequence that we analyzed
in Section \ref{II}. Let us change perspective. Let us explicity focus on
the events corresponding to changing sign. For this reason we build
up a new sequence where the sites of the laminar region are given the
vanishing value. This means that the states of constant velocity are
here perceived as states of rest. The random walker can make a jump
only at the moment when the time series analyzed in Section II changes
signs. 
 If at the moment of changing sign the walker makes a jump, of
the same intensity,
forward or backward, according to whether the sign change is positive
or negative, we shall refer to this process as {\em Symmetric Jump Model}
(SJM).
The scaling corresponding to the SJM has been studied by Shlesinger
in the pioneer paper of Ref.\cite{Shlesinger}, which yields the following
prescription when the condition $1<\mu<2$ applies,
\begin{equation}
\delta = \frac{1}{2}(\mu -1).
\label{zumofenandklafter}
\end{equation}
When $\mu>2$, the theory of Ref. \cite{Shlesinger} predicts
\begin{equation}
\delta=1/2.
\end{equation}

We see that the whole region $\mu > 2$, including that with $\mu < 3$,
would be indistinguishable from an ordinary Brownian diffusion. For
this reason, in addition to focusing our attention on the events
corresponding to the sign change, we decide to disregard whether the
change of sign is from - to + or vice-versa. This generates a kind of
asymmetric diffusion process that we term {\em Asymmetric Jump Model}
(AJM). This section is devoted to the discussion of the AJM by means
of two distinct theoretical approaches, both resulting in the same
prediction, an interesting form of asymmetric Levy diffusion. Then we
generate two artificial sequences corresponding to  $\mu > 2$ and $\mu <
2$, respectively, and we analyze them  by means of the DE method.

\subsection{Continuous Time Random Walk}\label{IIIA}

      The first theoretical approach rests on the Continuous
Time Random Walk (CTRW) of Montroll and Weiss \cite{montrollweiss}.
This means that
the distance between two nearest neighbor events, $\tau_{i} \equiv
t_{i+1} - t_{i}$, is considered to be a random number, described by
the probability density $\psi(\tau)$. Note that we are here using the
notation $\tau$ rather than $t$ to make clear to the reader the
difference between the absolute time t and the time distance between
two nearest neighbor events. As earlier said, we assume that the
random walker always moves in the same direction. Moreover, we also
set the condition that the steps have constant length.
Thus, the probability that a step of length $x$ is taken is given by
\begin{equation}
\label{constantstep}
\Pi(x) = \delta(x-W).
\end{equation}
Applying to this case the CTRW formalism of Montroll and
Weiss\cite{montrollweiss}, we write for $p(x,t)$, the
probability of the walker being at position $x$ at time $t$,
\begin{equation}
p(x,t) = F^{-1}L^{-1} \hat{p}(k,s) \equiv F^{-1}L^{-1}[ 
\frac{1-\hat{\psi}}{s}
\frac{1}{1-\lambda(k)\hat{\psi}(s)}],
\label{mainpredictionofweissandmontroll}
\end{equation}
where $\lambda(k)$ and $\hat{\psi}(s)$ are the Fourier and Laplace
transform of $\Pi(x)$ and $\psi(t)$, respectively. Note that
\begin{equation}
\lambda(k) = exp(ikW).
\label{explicitexpression}
\end{equation}
By Taylor series expansion of
Eq.(\ref{mainpredictionofweissandmontroll}) we get:
\begin{equation}
\label{intermediatestep}
\hat{p}(k,s) =\frac{1-\hat{\psi}}{s} \sum_{n =0}^{\infty}
\hat{\psi}^{n} e^{inkW}
\end{equation}
By evaluating the inverse Fourier transform of
Eq.(\ref{intermediatestep}) we arrive at:
\begin{equation}
\label{almosthere}
\hat{p}(x,s) =\frac{1-\hat{\psi}}{s} \sum_{n =0}^{\infty}
\hat{\psi}^{n} \delta(x - nW).
\end{equation}
This means that for large distances we can adopt the following
expression
\begin{equation}
\label{partialinversion}
p(x,s) = [\hat{\psi}(s)]^{\frac{x}{W}} \frac{1-\hat{\psi}(s)}{s}.
\end{equation}

We shed light on the meaning of Eq.(\ref{partialinversion})
with the following remarks. Let us denote by $\psi_{x}(t)$ the
probability of making $x$ jumps up to time $t$, with the last
jump occurring at $t$. Let us denote by $\Phi_{n}(t)$ the
probability of making $n$ jumps at earlier times, $t' < t$. The
two functions are connected by the relation
\begin{equation}
\Phi_{n}(t) = \int_{0}^{t}\psi_{n}(t')\Psi(t'-t)dt',
\label{keyconnection}
\end{equation}
where
\begin{equation}
\Psi(t) \equiv \int_{t}^{\infty}\psi(t') dt'.
\label{survivalprobability}
\end{equation}
Using the properties $\hat{\psi}_{n}(s) = [\hat{\psi}(s)]^{n}$ and
$\hat{\Psi}(s) = (1 -\hat{\psi}(s))/s$, and Laplace transforming
Eq.(\ref{keyconnection}), we get
\begin{equation}
\hat{\Phi}_{n}(s) = \hat{\psi}_{n}(s) \hat{\Phi}(s)
= [\hat{\psi}(s)]^{n} \frac{1-\hat{\psi}(s)}{s}.
\label{givingmeaning}
\end{equation}
We see that the result of Eq.(\ref{partialinversion}) can be
recovered by identifying $n$ with $x/W$, as it is legitimate to do
when the number of jumps done is so large as to perceive $n$ as a
continuous number.

Let us focus now our attention on the case where the distribution of
waiting times, $\psi(t)$, has the following asymptotic behavior
\begin{equation}
\psi(t) \approx \frac{const}{t^{\mu}},
\label{inversepowerlaw}
\end{equation}
with $\mu >1$ so as to insure the normalization condition. Let us
focus our attention on the case where
\begin{equation}
1 < \mu < 2.
\label{subdiffusion}
\end{equation}

This is the region where jumps of intensity $x$
occurring with equal probability in both the positive and the
negative direction of the $x$-axis, i.e. the SJM, would result\cite{zumofen}
in a diffusion
process with the scaling $\delta$ of Eq.(\ref{zumofenandklafter}).
We now show that, as  a benefit of adopting the AJM rather than the
SJM, we derive  an enhanced scaling parameter.
In fact,
for very small values of $s$,
\begin{equation}
\hat{\psi}(s) \approx 1 - c s^{\gamma},
\label{longtimelimit}
\end{equation}
where
\begin{equation}
\gamma = \mu - 1.
\label{definitionofgamma}
\end{equation}
This is the same property as that used by Zumofen and
Klafter\cite{zumofen}. By plugging 
Eq.(\ref{longtimelimit}) into
Eq.(\ref{partialinversion}) and by inverting the 
Laplace transform
with the nice method of ref.\cite{barkai}, we obtain
\begin{equation}
\label{asymmetriclevy}
p(x,t) \approx \frac{t}{x^{\frac{1}{\gamma} + 1}}
L(\frac{t}{x^{\frac{1}{\gamma}}},\gamma,-\gamma),
\end{equation}
where $L(y,\gamma,-\gamma)$ denotes a fully asymmetric L\'{e}vy stable
law of index $\gamma$\cite{levy}. It is straightforward to express
this important result in the form of Eq.(\ref{generalizedscaling})
with the time independent scaling
\begin{equation}
\delta = \mu - 1.
\label{enhanced}
\end{equation}
By comparing Eq.(\ref{enhanced}) to Eq.(\ref{zumofenandklafter})
it becomes evident the scaling enhancement produced by making the
diffusion process asymmetric. In fact, at $\mu = 3/2$ a
transition is made from subdiffusion to superdiffusion.

\subsection{Generalized Central Limit Theorem}
\label{IIIB}

The result of Section \ref{IIIA} can be recovered from within the attractive
perspective of the Generalized Central Limit Theorem
(GCLT)\cite{gnedenkokolmogorov} in a form derived from  properly
adopting the procedure of
Feller \cite{feller} to our purposes. To make closer the connection with the
formalism of
Feller let us assume $W = 1$. The function $p(x,t)$ in this
case is nothing but the probability for the walker to make $x$ steps
up to time $t$, or, equivalently, the probability of occurrence of
$x$ events up to time $t$. It has to be stressed that according to
the perspective expressed by Refs.\cite{mauro,massi}, in accordance
with the arguments of Gaspard and Wang\cite{gaspard}, this way of
proceeding can also be used to measure the complexity or randomness
of the process under study, in the special case where the time series
under study is totally memory-less. In fact, in this specific case the
occurrence of any sporadic event does not have any connection with
the earlier or subsequent events. It has to be pointed out that
the kind of time series that we are having in mind in this paper
is characterized by events of the same intensity, or made so by
ignoring the real intensity of each event. If there is no correlation
between intensity and time occurrence of the event, the simplifying
assumption that the events have the same intensity does not have any
significant consequence on the scaling of the resulting diffusion.

If the simplifying assumption applies, that no form of memory exists,
in addition to the inverse power law nature of the waiting time
distribution $\psi(t)$, we can safely apply the Feller arguments.
Feller \cite{feller}  studied the case where the survival
probability $\Psi(t)$ has the following time asymptotic expression
\begin{equation}
\label{asymptoticsurvivalprobability}
\Psi(t) \approx \frac{1}{t^{\gamma}}.
\end{equation}
The Feller notation is related to our notation by $\gamma = \mu -1$. In
the article of  Ref.\cite{feller}, Feller studies bot the case $0 < \gamma < 1$
and the case $1 < \gamma < 2$, which refers to our two regimes $1 <\mu <
2$ and $2 < \gamma < 3$, respectively. In the former case, Feller finds
the important property
\begin{equation}
\label{GIAK1}
Pr \left ( N_t \geq w t^{\gamma} \right )= G_{\gamma}(w^{-1/\gamma}),
\end{equation}
where, according to the notation used by Feller \cite{feller} $Pr(N_{t} > y)$
denotes the probability that the number of events occurring at times
$t^{\prime} \leq t$ is greater than $y$.
The function $G_{\gamma}(x)$ denotes the stable law whose
characteristics function is given by
\begin{equation}
\hat G_{\gamma}(z)=\exp \left \{ -|z|^{\gamma} \left [
\cos \left ( \frac{\pi \gamma}{2} \right ) -
 i \sin \left (\frac{\pi \gamma}{2} \right )
sgn(z) \right ] \Gamma (1 - \gamma) \right \}.
\end{equation}
The symbol $\Gamma$ denotes the Euler function. The meaning of the
function $G_{\gamma}(w)$ is made transparent by the case where the
walker makes jumps in the same direction, with the probability
$\Pi(x) \propto 1/x^{\gamma}$, at regular interval of times. In this specific
case, which is the case of an asymmetric Levy flight, the position of
the random walker at time $t$, $X_{t}$,  fulfills the property
\begin{equation}
\label{trentasette}
Pr \left ( X_t \leq w t^{1/\gamma} \right )= G_{\gamma}(w).
\end{equation}
The function on the right hand side of Eq. (\ref{trentasette}) is the same as that
appearing in the right hand side of Eq. (\ref{GIAK1}). However, 
we know that it has to do with an asymmetric Levy process. 
Consequently, it is a stable law that the theory of \cite{levy} 
makes it possible to write under the form
\begin{equation}
G_{\gamma}(w)=\int \limits_{0}^{w} L(y,\gamma,-\gamma) {\rm d}y,
\end{equation}
where $L(y,\gamma,-\gamma)$ is the function defined in Section \ref{IIIA}.
We note that Eq. (\ref{GIAK1}) can also be written as
\begin{equation}\label{grigo}
\int_{w t^{\gamma}}^{\infty} {\rm d}N_{t} p(N_{t}) = G_{\gamma}(w^{\gamma}) .
\end{equation}
By differentiating Eq. (\ref{grigo}) with respect to $w$ and 
setting$N_{t}=x$ (the position of the walker at time $t$,
labeled as $x$, is equal to the number of steps taken up to 
time $t$, namely, $N_{t}$) we get
\begin{equation}
p(x,t) \approx \frac{t}{x^{\frac{1}{\gamma} + 1}}
L(\frac{t}{x^{\frac{1}{\gamma}}},\gamma,-\gamma),
\label{feller1}
\end{equation}
which agrees with the scaling predicted by the CTRW arguments 
used in section \ref{IIIA} since it leads to the scaling
parameter $\delta = \mu - 1$.

In the case where $1 < \gamma < 2$ we are led by the 
corresponding Feller prescription \cite{feller} to express the
probability distribution in the reference translating with the 
mean velocity (the inverse of the mean waiting time). With an 
argument similar to the one earlier used we obtain
\begin{equation}
p(x,t) \approx \frac{1}{t^{\frac{1}{\gamma} }}
L(\frac{x}{t^{\frac{1}{\gamma}}},\gamma,-\gamma),
\label{feller2}
\end{equation}
and it is now easy to prove that this yields, for $2 < \mu < 3$, the 
scaling prescription
\begin{equation}
\delta = \frac{1}{\mu - 1}.
\label{levyscalingagain}
\end{equation}
Results are summarized in Tab. 1.

\subsection{Numerical Calculations}

\begin{figure}[h]
\begin{center}
\includegraphics[scale=.6]{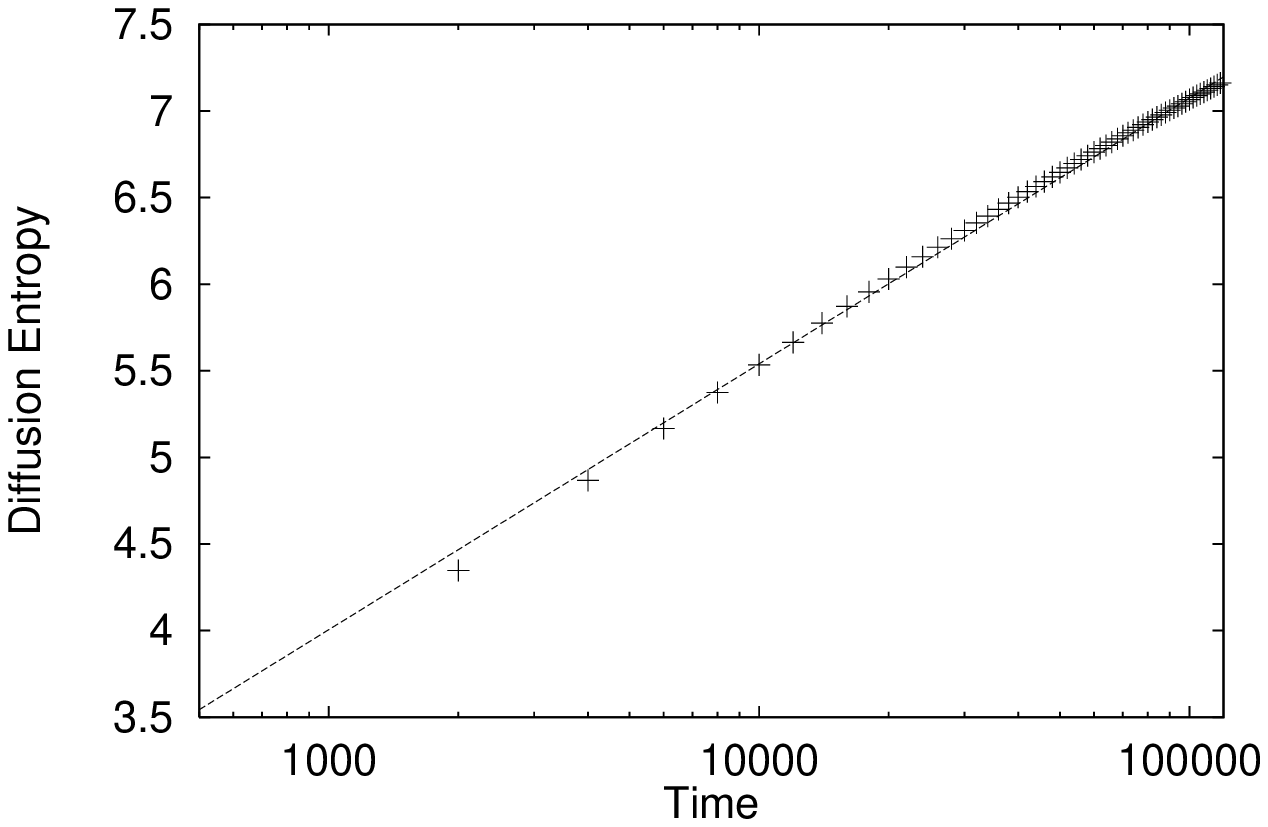}
\caption{\label{AJMDE}
The diffusion entropy as a function of time. The numerical method is
applied to the artificial sequence of Section \ref{II}, with $\mu = 2.5$,
studied according to the AJM prescription. According to the
theoretical arguments of the text the scaling parameter $\delta$ is the
slope of the straight line fitting the numerical results at large
times, which yields in this case $\delta = 2/3 = 1/(\mu -1)$ ($T = 10$).}
\end{center}
\end{figure}

This section is devoted to illustrating the results of numerical
calculations that confirm the benefits of the strategy suggested
in this paper.
First of all with Fig.\ref{AJMDE} we show that the AJM makes it
possible to reach much earlier the correct L\'{e}vy scaling 
of Eq.(\ref{levyscaling}). The numerical work shows that the time scale
at which the DE method yields the correct scaling is, in this case,
equal to about $2,000$ ($T=10$).

\begin{figure}[h]
\begin{center}
\includegraphics[scale=.6]{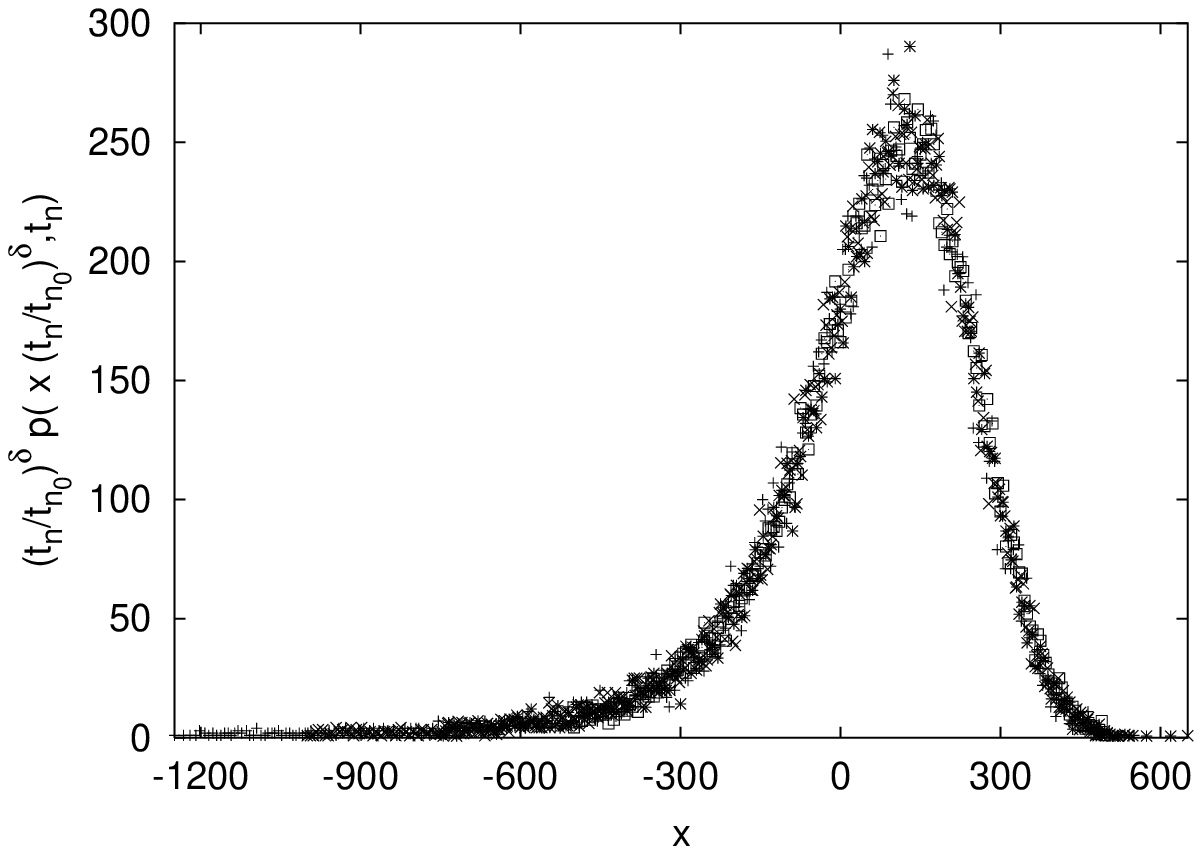}
\caption{\label{scaling25}
 Scaled distributions as a function of $x$. These scaled
distributions refer  to the artificial sequence of Section \ref{II}, with
$\mu = 2.5$, dealt with according to the AJM prescription. The times considered
are: $t_n$=2500 '+', 2000 '$\star$', 1500 '$\times$',
1000 '$\boxdot$'. Note that $t_{n_{0}} = 1000$.
Time is expressed in units of T, with T defined by 
Eq.(\ref{chosenanalyticalform}), and T = 10.}
\end{center}
\end{figure}

We also prove that the resulting scaling value is a genuine reflection 
of the property of Eq.(\ref{rigorousdefinition}). To do so, we
express the diffusion process in the reference framework moving with
the velocity $TW/(\mu-2)$ corresponding to the theoretical
prediction of section \ref{IIIB}. Then we consider a reference time
$t_{n_{0}}$ and later times $t_{n} > t_{n_{0}}$. 
For any time $t_{n}$ we consider the squeezing parameter 
$R(n) = (t_{n}/t_{n_{0}})^{\delta}$, with $\delta$ given by the scaling 
detected by the DE method. The distance scale is squeezed by the 
quantity $R(n)$ and the ordinate scale is enhanced by the
quantity $1/R(n)$. If the scaling detected by the DE is correct and 
if this procedure is applied to the distribution corresponding 
to the time $t_{n}$ this scaling procedure should make it 
identical to the distribution corresponding to the time $t_{n_{0}}$. Fig. 
\ref{scaling25} shows that this is really so, since all the 
distributions at times $t_{n}$ are made to virtually coincide
with the distribution at time $t_{n_{0}}$.

\begin{figure}[h]
\begin{center}
\includegraphics[scale=.6]{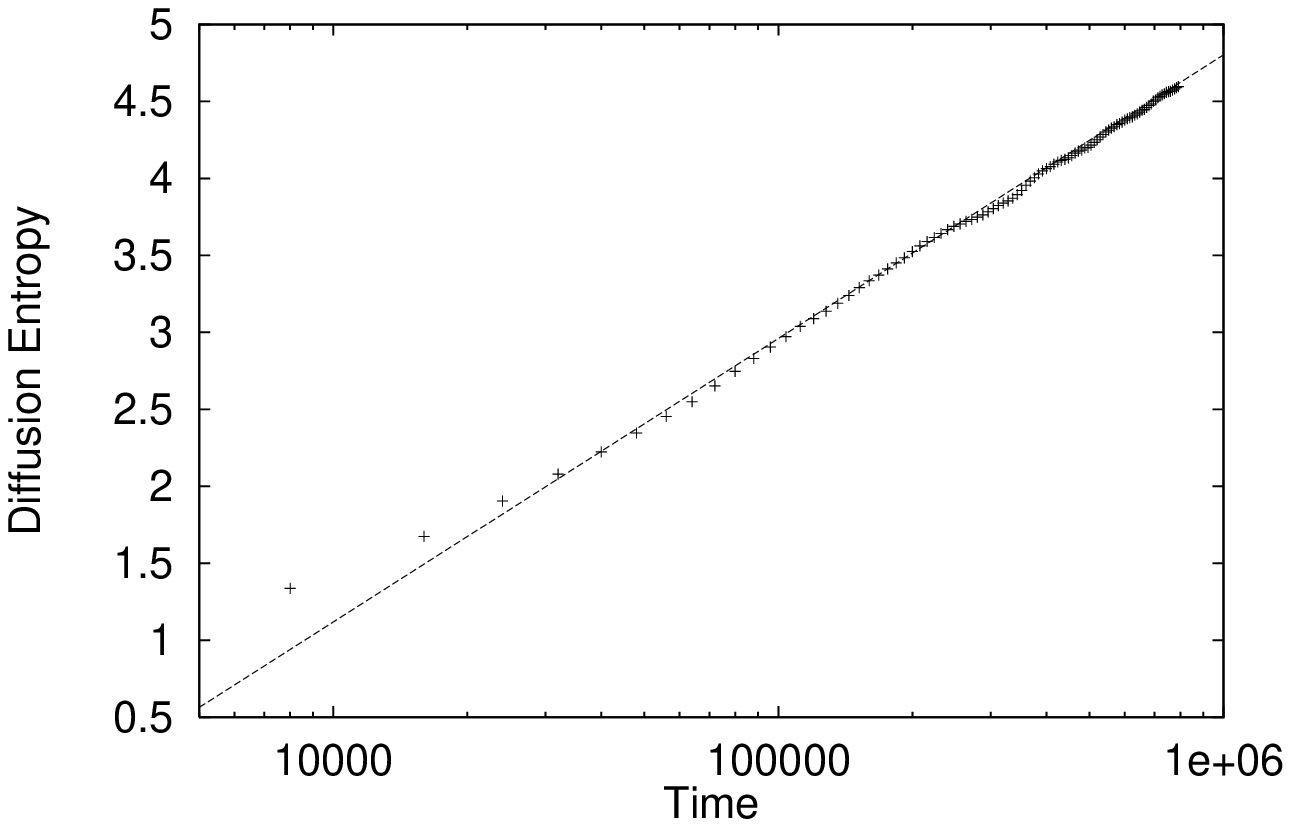}
\caption{\label{AJM18}
The diffusion entropy as a function of time. The numerical method is
applied to the artificial sequence of Section \ref{II}, with $\mu = 1.8$,
studied according to the AJM prescription. According to the
theoretical arguments of the text the scaling parameter $\delta$ is the
slope of the straight line fitting the numerical results at large
times, which yields in this case $\delta = \mu -1= 0.8$ ($T = 10$).}
\end{center}
\end{figure}

\begin{figure}[h]
\begin{center}
\includegraphics[scale=.6]{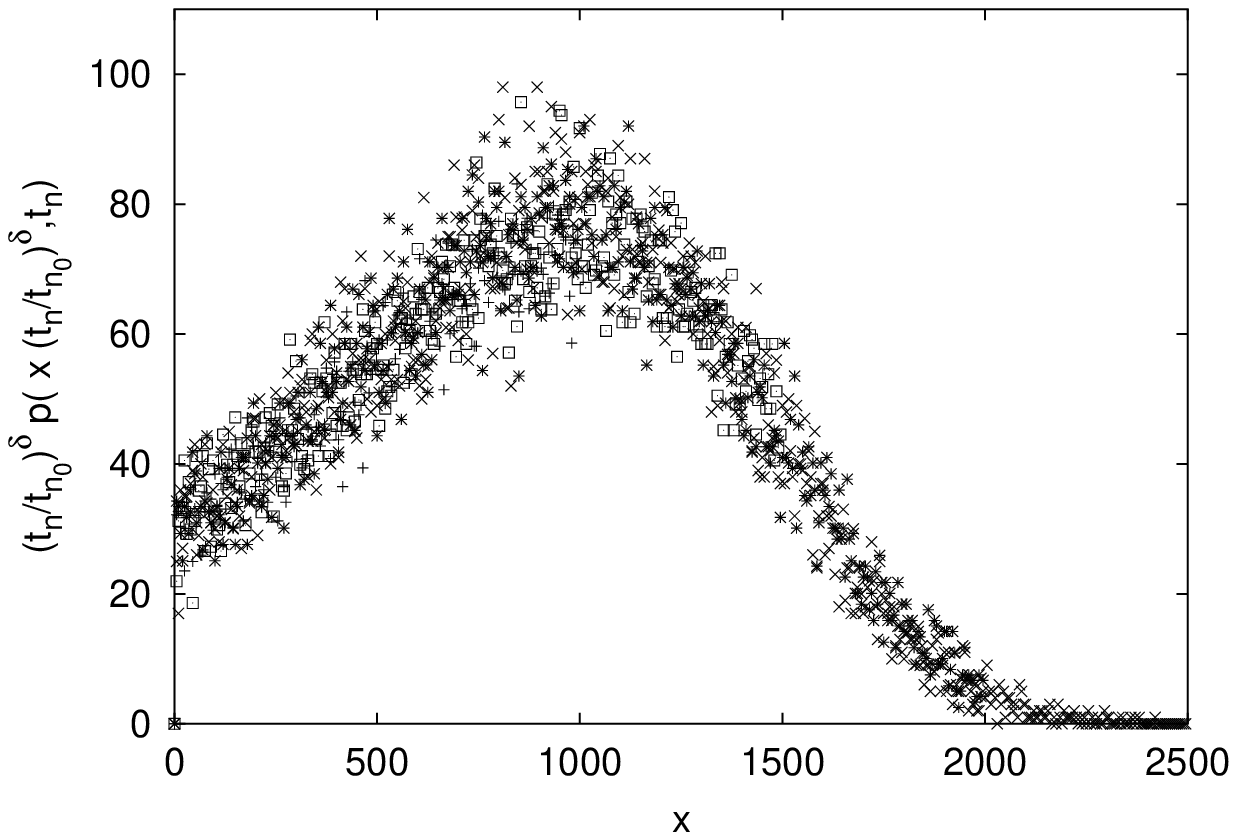}
\caption{\label{scaling18}
Scaled distributions as a function of $x$. These scaled
distributions refer  to the artificial sequence of Section \ref{II},
with $\mu = 1.8$, dealt with according to the AJM prescription.
The times considered
are: $t_{n}=$ 2500 '+', 2000 '$\star$', 1500 '$\times$', 1000 '$\boxdot$'.
Note that $t_{n_{0}} = 1000$.
Time is expressed in units of T, with T defined by
Eq. (\ref{chosenanalyticalform}), and $T = 10$.}
\end{center}
\end{figure}

We have explored also the case $\mu < 2$.
Figure \ref{AJM18} shows the result of the DE method in this case, and
Fig.\ref{scaling18} shows the result of scaling back to 
earlier times the distributions corresponding to later times, in this 
case. The result is not as accurate as that
illustrated by Fig.\ref{scaling25}. However, we judge it to be 
satisfactory, especially if we take into account the fact 
that the region $\mu < 2$ corresponds to the case of extremely 
large time correlations, with an infinite sojourn time.

\section{The Method in Action in the Case of Real Sequences}\label{cuore}

\begin{figure}[h]
\begin{center}
\includegraphics[scale=.6]{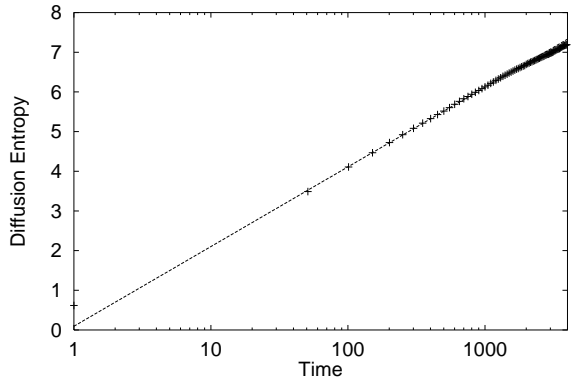}
\caption{\label{fig6}
The DE method applied to the heart beating sequence Ltdb15814,
changed into a diffusion process according to the AJM prescription.
The straight line is the best fit made with
Eq. (\ref{keyrelation}) and it yields $\delta=0.872$.}
\end{center}
\end{figure}

\begin{figure}[h]
\begin{center}
\includegraphics[scale=.6]{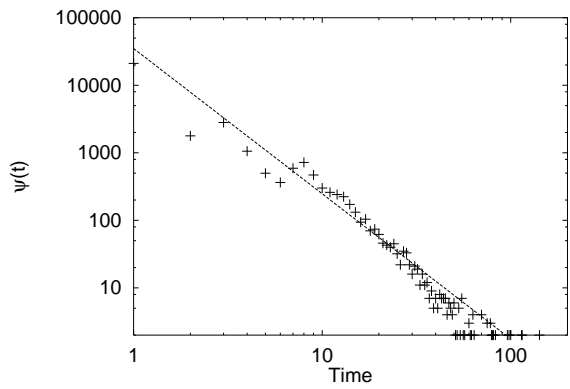}
\caption{\label{fig7} Coarse-grained waiting time distribution
$\psi(t)$ for Ltdb15814. The straight line is an inverse powe law
curve of the form $C/t^{\mu}$. According to the theory of 
section \ref{III}, $\mu=1+1/\delta$.} 
\end{center}
\end{figure}

As an example of how this technique can be applied
to real data we
considered time series concerning human heartbeats. 
These series were taken from Physionet \cite{physio}
 and are a record of time intervals between
successive beats. 
To apply our method we adopted a coarse graining
procedure: times between beats were first multiplied by
$10$, with only the integer part of the resulting number taken into account.
This sequence of integer numbers was then transformed into a
new one according to the following prescription: we wrote $0$ if the
number was equal to the preceeding one, $1$ if it was different. We note
that this would have resulted almost in a sequence of only 1's, without any 0,
 had we not adopted the coarse graining procedure. From this sequence we
generated an asymmetric diffusion process (AJM) which we then studied by means of
DE method described in the previous sections 
to derive the scaling parameter $\delta$. 
This corresponds to having a walker taking a step in a fixed direction when the
time interval between successive beats, as seen on a coarse grained scale,
differs from the preceding one.
 We also analyzed the probability distribution $\psi(t)$ of waiting times 
between the $1'$s, which shows an inverse power-law 
behavior with index $\mu > 2$.
We thus expect the correct scaling to be predicted by Eq.(\ref{levyscalingagain})
rather than by Eq.(\ref{enhanced}).
Results obtained from different databases are reported in table \ref{LTDB}, whose
third column represents the value of $\mu$ we derive from $\delta$. An example
for a particular database (Ltdb15814) is shown in Fig.\ref{fig6} where we derive the
scaling $\delta$ using DE, and in Fig.\ref{fig7}, where we show how the inverse power
law of index $\mu = 1 + 1/\delta$ indeed fits the data used to obtain $\psi(t)$.

\section{Concluding Remarks}\label{V}

 This paper is a sequel to that of Ref.\cite{nicola}, where the method
of DE was applied for the first time. However, the emphasis here is
on the detection of the scaling more than on the property that the
authors of Ref.\cite{nicola} referred to as memory. On the basis of the
results of the present paper, the memory of Ref.\cite{nicola} seems to have to
do with the transition from the short-time to the long-time regime.
The main result here reached, however, has to do with proving the
convenience of recording the times at which events occur, an event
being a sign change, regardless of whether it implies an increase or a
decrease of the intensity signal. If we record the time at which the
event occurs, and we interpret it as a jump of the random walker
always in the same direction, with a fixed intensity, then the
diffusion process becomes asymmetric, with the big benefit of
enhancing the value of the scaling parameter in the interval $1< \mu <
2$, and of making it sensitive to the power law index $\mu$ in the whole
interval $1 < \mu < 3$. The AJM has to be compared to the SJM, a model
that the theoretical analysis of the 1974 pioneer work of Shlesinger
\cite{Shlesinger} shows to be insensitive to $\mu$ in the range
$2< \mu < 3$.
The value of the scaling parameter $\delta$, stemming from the AJM, as earlier
repeatedly pointed out, is enhanced with respect to that of the SJM,
in the whole interval $2< \mu < 3$. If we compare the AJM to the SVM of
Ref.\cite{zumofen}, we find that it provides information on $\mu$ even in the
interval $1 < \mu < 2$, where the SVM would steadily result in the
ballistic prescription $\delta = 1$.

       It is interesting to notice that the AJM proposed by this paper
emphasizes the phase transition character of a transformation moving
from $\mu > 2$ to $\mu < 2$. As pointed out in Ref.\cite{massi}, this is a
transition from a stationary to a non-stationary regime. In fact,
Ref. \cite{massi} shows that $\mu < 2$ implies the lack of an invariant
distribution, thereby making any observation carried out for a finite
time equivalent to monitoring an out of equilibrium process. We are
convinced that the systematic application of the prescriptions of
this paper to studying time series generated by complex dynamic
processes, might lead to discover these non-stationary cases.

        The application to the real sequences of physiological
interest of Section \ref{cuore} shows that the detection of the scaling
parameter $\delta$ is much more accurate than the direct evaluation of
$\mu$ from the experimental data, the only ambiguity being the fact
that the value of $\delta$ is compatible with both
$\mu(1) = 1 + 1/\delta$ and $\mu(2)  = 1 + \delta$. So, the experimental
error affecting the direct determination of delta must only be
smaller than the difference $\mu(1) - \mu(2)$. Only when the value of
$\mu$ is very close to the transition value $\mu = 2$, the adoption of
the DE method to detect $\mu$ becomes ambiguous, and additional
criteria must be found.

A possible efficient criterion could be given by the joint use of the SVM
and AJM method. In fact, as clearly illustrated by Table \ref{scalingriassunti},
 if the SVM yields $\delta = 1$, this means that $\mu < 2$.
 If, on the contrary, the SVM,
within the statistical accuracy of the numerical procedure, yields the
same scaling as the SVM, we have to select $\mu > 2$.

         Another benefit of the adoption of the AJM is the faster
attainment of the correct scaling. It is not yet quite clear which is
the true reason why the correct scaling of (\ref{levyscaling}),
in the region $2 <\mu< 3$, is significantly faster than with the SVM.
One reason might be that the probability distribution of the AJM
diffusion process is
sharper than that of the SVM.
Some more research work is required to assess this point. We hope that the
results already obtained are interesting enough as to draw the
attention of the researches of this field of investigation on the
benefit stemming from converting the experimental data into
asymmetric diffusion processes. As the last but not least remark, we
want to point out that the DE method is probably the most convenient
way to detect the scaling of an asymmetric diffusion processes, since
this technique is not affected  by the presence of biases, and does
not require the adoption of any form of de-trending.

\newpage

\begin{table}[!h]
\centering
\begin{tabular}{|cc|}
\hline
\multicolumn{2}{|c|}{SVM}\\
$1<\mu<2$ & $\delta=1$ \\
$2<\mu<3$ & $\delta=1/(\mu-1)$\\
\hline
\multicolumn{2}{|c|}{SJM} \\
$1<\mu<2$ & $\delta=(\mu-1)/2$\\
$2<\mu<3$ & $\delta=1/2$ \\
\hline
\multicolumn{2}{|c|}{AJM}  \\
$1<\mu<2$ & $\delta=\mu-1$ \\
$2<\mu<3$ & $\delta=1/(\mu-1)$ \\
\hline
\end{tabular}
\caption{\label{scalingriassunti} The scaling parameter $\delta$
as a function of $\mu$ for SVM, SJM and AJM.
The AJM is the only model where $\delta$ always
changes upon changing $\mu$.}
\end{table}

\begin{table}[!h]
\label{LTDB}
\centering
\begin{tabular}{|c|c|c|}
\hline
{\bf Archive} & $\mathbf{\delta}$ & $\mathbf{\mu=1+1/\delta}$ \\
\hline
14046 & $0.777 \pm 0.002$ & $2.29$ \\
14134 & $0.863 \pm 0.006$ & $2.16$ \\
14172 & $0.876 \pm 0.002$ & $2.14$ \\
14184 & $0.804 \pm 0.005$ & $2.24$ \\
14157 & $0.855 \pm 0.010$ & $2.17$ \\
15814 & $0.872 \pm 0.002$ & $2.15$ \\
\hline
\end{tabular}
\caption{Values of the scaling parameter $\delta$
obtained with the DE method.
The archive numbers correspond to the denomination of
MIT-BIH Long-Term ECG Database
as reported in ref.\cite{physio}.}
\end{table}

\end{document}